\documentclass[prb,twocolumn,a4paper]{revtex4}
\usepackage{amsfonts}
\usepackage[intlimits]{amsmath}
\usepackage{amssymb}
\usepackage[dvipdf]{graphicx}

\begin{document}

\title{Spin-polarized tunneling current through a ferromagnetic insulator between two metallic or superconducting leads}
\author{N. Sandschneider}
\email{niko.sandschneider@physik.hu-berlin.de}
\author{W. Nolting}
\affiliation{Festk\"orpertheorie, Institut f\"ur Physik, Humboldt-Universit\"at zu Berlin, Newtonstr. 15, 12489 Berlin, Germany}

\begin{abstract}
Using the Keldysh formalism the tunneling current through a hybrid structure where a confined magnetic insulator (I) is sandwiched between two non-magnetic leads is calculated. The leads can be either normal metals (M) or superconductors (S). Each region is modelled as a single band in tight-binding approximation in order to understand the formation of the tunneling current as clearly as possible. The tunneling process itself is simulated by a hybridization between the lead and insulator conduction bands. The insulator is assumed to have localized moments which can interact with the tunneling electrons. This is described by the Kondo Lattice Model (KLM) and treated within an interpolating self-energy approach. For the superconductor the mean-field BCS theory is used. The spin polarization of the current shows a strong dependence both on the applied voltage and the properties of the materials. Even for this idealized three band model there is a qualitative agreement with experiment.
\end{abstract}

\maketitle

\section{Introduction}
The understanding of spin-polarized tunneling currents in mesoscopic hybrid structures is a very active field in current research \cite{Prinz}\cite{Moodera99}\cite{Tsymbal}. This interest stems from the possibility to build electronic devices which take into account the spin degree of freedom of the electrons ("spintronics"). Usually magnetic tunnel junctions (MTJs) consist of two ferromagnetic metals separated by a thin non-magnetic insulator. However, in this paper we consider a different geometric structure. The two outer leads shall be either normal metals (M) or superconductors (S) while the central region is always assumed to be a ferromagnetic insulator. The band splitting in the insulator due to the exchange interaction will lead to different tunnel barrier heights for spin-up and spin-down electrons, resulting in a finite spin polarization of the tunneling current. This is called the spin-filter effect.\\
Since a MTJ is build by coupling materials with in general different chemical potentials, it is inherently out of thermal equilibrium. Therefore the usual many-body description has to be modified. Its extension to non-equilibrium phenomena is the so-called Keldysh formalism \cite{Keldysh}.\\
Pioneering work in this field has been done by Caroli et al.\cite{Caroli} who first presented a model Hamiltonian for calculating the tunneling current. The model was later refined by Meir and Wingreen\cite{Meir}. Heide et al.\cite{Heide} investigated the relation between the current and the interlayer exchange coupling. A similar model to the one presented in this paper was used by Zeng et al.\cite{Zeng} to calculate transport properties in hybrid structures. They use ferromagnetic and superconducting leads with a non-magnetic central region while in our case the situation is reversed.\\
Most experiments on spin filtering were done with Europium chalkogenide barriers and metallic contacts. Recently there also have been several experiments where other materials were used. Figielski et al.\cite{Figielski} performed experiments for EuS barriers embedded into a semiconducting PbS matrix. Gajek et al.\cite{Gajek} used $\text{BiMnO}_3$ as tunnel barrier and the half-metallic oxide $\text{La}_{2/3}\text{Sr}_{1/3}\text{O}_3$ as counter-electrode.\\
The paper is organized as follows: in section \ref{Sec:Theory} we develop the theoretical model and derive a general formula for the tunneling current. In sections \ref{Sec:MIM} and \ref{Sec:MIS} the model is evaluated for the M/I/M- and M/I/S-systems, respectivly. Special emphasis is made to understand the current and spin polarization features in terms of the quasiparticle density of states (QDOS). In the last section \ref{Sec:Conclusion} we shortly summarize the paper and give an outlook with possible improvements and extensions of the model.

\section{Model and Theory}\label{Sec:Theory}
In this section we formulate a model Hamiltonian which will allow us to simulate the tunneling of electrons through an insulating region. The Hamiltonian was first used by Caroli et al. \cite{Caroli}. In Fig. \ref{fig:princ} the geometric setup is shown schematically. At the beginning, i.e. at $t=-\infty$, it is assumed that there is no contact between the three regions. Therefore each one is in thermal equilibrium, characterized by its respective chemical potential. Then the coupling between them is turned on, resulting in an overall chemical potential in the whole system. This coupling will be described by perturbation theory in non-equilibrium. We assume the tunneling to be ballistic, i.e. there is no scattering of electrons involved. Hence the tunneling is elastic and the energy of the electrons is conserved. Furthermore we assume translation invariance in each region.\\
The model Hamiltonian consists of three parts:
\begin{figure}
\includegraphics[width=\linewidth]{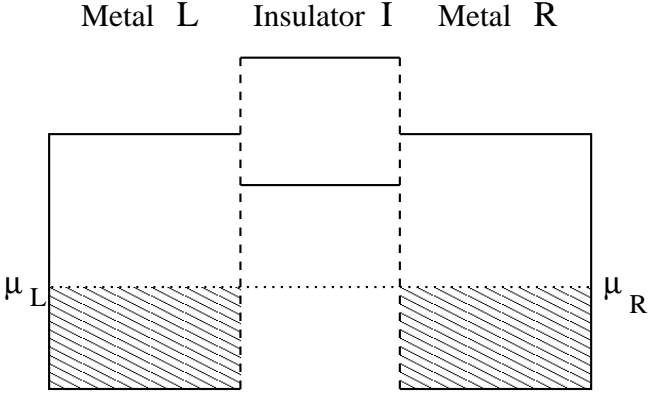}
\caption{Tunnel junction without applied voltage for $T=0$. The rectangles symbolize the conduction bands of each layer. The chemical potential is equal for all three layers and marked by a dotted line. Occupied states in the metals are hatched.}\label{fig:princ}
\end{figure}
\begin{equation}\label{form:ModelH}
H=H_L+H_{ins}+H_T
\end{equation}
$H_L$ describes the leads, $H_{ins}$ the insulating region and $H_T$ the tunneling between them. For now the contacts are assumed to be non-interacting, thus
\begin{equation}
H_L = \sum_{\substack{M \\ M=L,R}} \sum_{\mathbf{k}_M\sigma} \epsilon_{\mathbf{k}_M} c_{\mathbf{k}_M\sigma}^+ c_{\mathbf{k}_M\sigma}
\end{equation}
$c_{\mathbf{k}_M\sigma}$ ($c_{\mathbf{k}_M\sigma}^+$) is the annihilation (creation) operator which annihilates (creates) an electron with spin $\sigma$ and wave vector $\mathbf{k}_M$ in lead $M$ where $M=L$ for the left lead and $M=R$ for the right lead. $\epsilon_{\mathbf{k}_M}=\frac{1}{N_M}\sum_{i_Mj_M}t_{i_Mj_M}e^{-i\mathbf{k}_M\cdot(\mathbf{R}_{i_M}-\mathbf{R}_{j_M})}$ is the Fourier transformed hopping integral between lattice sites $\mathbf{R}_{i_M}$ and $\mathbf{R}_{j_M}$. In the case of superconducting leads $H_L$ has to be modified. This will be discussed in section \ref{Sec:MIS} in detail.\\
For the derivation of the tunneling current formula, it is not yet necessary to specify the interaction $H_{int}$ in the insulator. The only restriction is that it has to commute with the particle number operator of the leads. Since $H_{int}$ will in general be a functional of the insulator construction operators, i.e. $H_{int}=H_{int}(d_{\mathbf{k}_I\sigma,n}^+;d_{\mathbf{k}_I\sigma,m})$, this is true for all usual interactions. Hence
\begin{equation}
H_{ins} = \sum_{\substack{\mathbf{k}_I\sigma \\ n,m}} \epsilon_{\mathbf{k}_I}^{nm} d_{\mathbf{k}_I\sigma,n}^+ d_{\mathbf{k}_I\sigma,m} + H_{int}(d_{\mathbf{k}_I\sigma,n}^+;d_{\mathbf{k}_I\sigma,m})
\end{equation}
The construction operators in the insulator are labeled by the letter $d$. $H_{ins}$ is written in a multi-layer form for reasons that will be discussed later on. So $d_{\mathbf{k}_I\sigma,n}$ ($d_{\mathbf{k}_I\sigma,n}^+$) destroys (creates) an electron with spin $\sigma$ and wave vector $\mathbf{k}_I$ in the layer $n$ of the insulator.\\
It remains the definition of the central part of the model, namely the tunneling Hamiltonian $H_T$. Tunneling is modelled by a hybridization between the conduction bands of the leads with the (empty) conduction band of the insulator. Therefore we choose the following ansatz:
\begin{equation}\label{form:HT}
H_T = \sum_{\substack{M \\ M=L,R}}\sum_{\substack{\mathbf{k}_M \mathbf{k}_I\sigma\\ n}} \left(\epsilon_{\mathbf{k}_M \mathbf{k}_I}^n c_{\mathbf{k}_M\sigma}^+ d_{\mathbf{k}_I\sigma,n} + H.c.\right)
\end{equation}
\begin{figure}
\includegraphics[width=\linewidth]{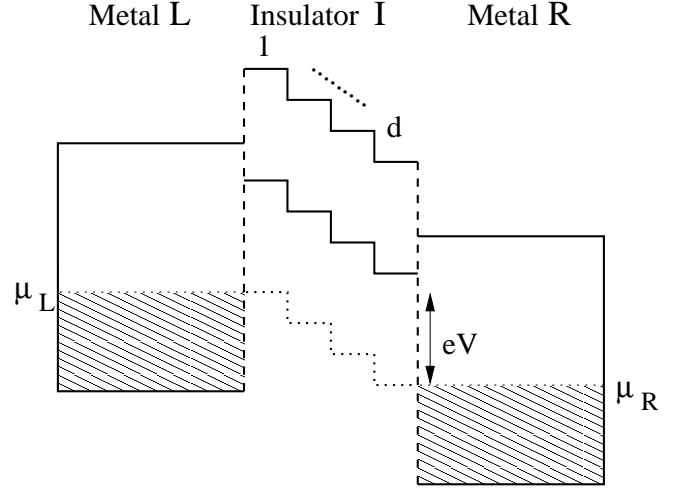}
\caption{Simulation of the tilting of the insulator band by d layers with shifted band centers. They are shifted according to Eq. (\ref{equ:vshift}). Occupied states are hatched and the chemical potential is shown as a dotted line.}\label{fig:hv}
\end{figure}
H.c. is the Hermitian conjugate of the first term. The coupling constants $\epsilon_{\mathbf{k}_M \mathbf{k}_I}^n$ are parameters which determine the strength of the hybridization between the bands. Their value is initially unknown and it is not possible to determine it within the model. However, they can be fixed by comparison with experiment.\\
Still the Hamiltonian is not complete, since the voltage $V$ does not appear anywhere. It will lead to a rigid shift of $e\cdot V$ between the band centers of the leads and the insulator band will be tilted accordingly. The tilting of the insulator conduction band cannot be modelled by a single insulator layer, since it can only have one center of gravity. Therefore the insulator is replaced by several insulating layers\cite{Metzke} whose band centers are shifted according to Fig. \ref{fig:hv}. This is described by the additional Hamiltonian
\begin{equation}
H_V = \sum_{\substack{i_M \sigma \\ M=L,R}} V_M c_{i_M\sigma}^+ c_{i_M\sigma} + \sum_{\substack{i_I\sigma \\ n}} V_I^n d_{i_I\sigma,n}^+ d_{i_I\sigma,n}
\end{equation}
The $V_M$ and the $V_I^n$ are the potentials in the leads and the insulating layers, respectively. Their relation to the applied voltage $V$ is as follows
\begin{eqnarray}
V_L &=& 0 \nonumber\\
V_R&=& V \label{equ:vshift}\\
V_I^n&=&V\frac{n-1}{d-1}\quad(n=1,\dots,d)\nonumber
\end{eqnarray}
This choice will lead to the behavior shown in Fig. \ref{fig:hv}. Since $H_V$ only changes the one-particle energies in each region by a fixed amount, it can be incorporated in the model Hamiltonian (\ref{form:ModelH}).\\
It is convenient to decompose the current flowing through the junction into two currents, one from the left ($J_L^\sigma$) and one from the right lead ($J_R^\sigma$). Since we are only interested in steady state currents, the symmetry relation $J_L^\sigma=-J_R^\sigma$ holds and the total current $J^\sigma$ can therefore be symmetrized:
\begin{equation}
J^\sigma = J_L^\sigma = \frac{1}{2}(J_L^\sigma+J_L^\sigma) = \frac{1}{2}(J_L^\sigma-J_R^\sigma)
\end{equation}
Deriving a formula for $J_L^\sigma$ will be the next task. It will be proportional to the change of electron number in the left lead:
\begin{equation}\label{form:ansatzJL}
J_L^\sigma(t) = -e\langle\dot{N}_L^\sigma(t)\rangle = \frac{ie}{\hbar}\langle[N_L^\sigma(t),H]_-\rangle
\end{equation}
The time dependence of the current is of a strictly formal nature. The commutator on the r.h.s. has to be evaluated to get a final expression for $J_L^\sigma$. It is basically identical to the lesser Green's function $G^<_{\mathbf{k}_I\mathbf{k}_L\sigma,n}(t,t')=i\langle c_{\mathbf{k}_L\sigma}^+(t')d_{\mathbf{k}_I\sigma,n}(t)\rangle$. This function can be calculated within the Keldysh formalism by using the analytic continuation rules by Langreth\cite{Langreth} on the corresponding non-equilibrium Green's function (NEGF) $G_{\mathbf{k}_I\mathbf{k}_L\sigma,n}(t,t')$. The details of the calculation are carried out in appendix \ref{sec:appcurr}. One gets
\begin{eqnarray}
J_L^\sigma &=& \frac{2e}{\hbar} \sum_{\substack{\mathbf{k}_L\mathbf{k}_I\\ n,m}} \mbox{Re}\Bigl(\int\frac{dE}{2\pi\hbar}\epsilon_{\mathbf{k}_L\mathbf{k}_I}^n \epsilon_{\mathbf{k}_L\mathbf{k}_I}^{m*}\cdot\label{form:resJL}\\ &\cdot&\Bigl(G_{\mathbf{k}_I\sigma,nm}^r(E) g_{\mathbf{k}_L}^<(E) + G_{\mathbf{k}_I\sigma,nm}^<(E) g_{\mathbf{k}_L}^a(E)\Bigr)\Bigr)\nonumber
\end{eqnarray}
The free metal Green's functions are given by
\begin{eqnarray}
g_{\mathbf{k}_L}^<(E) &=& 2\pi\hbar i f_L(\epsilon_{\mathbf{k}_L})\delta(E-\epsilon_{\mathbf{k}_L})\\
g_{\mathbf{k}_L}^a(E) &=& \frac{\hbar}{E-\epsilon_{\mathbf{k}_L}-i0^+}
\end{eqnarray}
where $f_L(E)$ is the Fermi function of the left lead. The insulator Green's functions $G_{\mathbf{k}_I\sigma,nm}^r(E)=\langle\langle d_{\mathbf{k}_I\sigma,n};d_{\mathbf{k}_I\sigma,m}^+\rangle\rangle$ and $G_{\mathbf{k}_I\sigma,nm}^<(E)=i\langle d_{\mathbf{k}_I\sigma,m}^+ d_{\mathbf{k}_I\sigma,n}\rangle$ can be calculated by using the equation-of-motion method. This is done in appendix \ref{sec:appG}. At this point of the calculation it is necessary to specify the interaction $H_{int}$ in the insulator, which is assumed to consist of localized moments. Since there are also itinerant electrons (through tunneling) in the insulator, the KLM is considered to be an appropriate model for this region. Thus
\begin{equation}
H_{int} = H_{sf} = -\frac{1}{2}J\sum_{\substack{i_I\sigma\\ n}}(z_\sigma S_{i_I,n}^z n_{i_I\sigma,n} + S_{i_I,n}^\sigma d_{i_I-\sigma,n}^+ d_{i_I\sigma,n})
\end{equation}
where $J$ is the exchange coupling constant and $\mathbf{S}_{i_I,n}$ denotes the spin operator of the localized moments. The electrons interact only at one lattice site at a time with the localized moments. Therefore the summation is over one layer index only. The interaction causes no transitions between the layers in other words. The KLM defines a complicated many-body problem and can only be solved approximately for most cases of interest. Here we use an interpolating self-energy approach (ISA) \cite{ISA1}. The self-energy of the KLM is in general defined as
\begin{equation}\label{equ:KLMself-energy}
\langle\langle[H_{int},c_{\mathbf{k}_I\sigma}]_-;c_{\mathbf{k}_I\sigma}^+\rangle\rangle = \Sigma_{\mathbf{k}_I\sigma}(E) G_{\mathbf{k}_I\sigma}(E)
\end{equation}
The ISA self-energy is an \emph{exact solution} of several limiting cases of the model. It fulfills the zero-bandwidth limit, ferromagnetic saturation, second-order perturbation theory and the high energy expansion. The approximation lies in the assumption that it is valid for all intermediate cases, too. It was found in other works that it gives reliable results for the materials usually described by the KLM, namely manganites, magnetic semiconductors and local-moment metals\cite{Stier}\cite{Sharma}\cite{Kreissl}. The ISA self-energy is given by the following expression
\begin{eqnarray}
\Sigma_\sigma(E) &\equiv& \frac{1}{N_I}\sum_{\mathbf{k}_I}\Sigma_{\mathbf{k}_I\sigma}(E) \nonumber\\
&=& -\frac{1}{2} J m_\sigma + \frac{1}{4} J^2 \frac{a_\sigma G_0(E-\frac{1}{2}J m_\sigma)}{1-\frac{1}{2}J G_0(E-\frac{1}{2}J m_\sigma)}
\end{eqnarray}
where
\begin{eqnarray}
G_0(E) &=& \frac{1}{N_I}\sum_{\mathbf{k}_I}\frac{1}{E-\epsilon_{\mathbf{k}_I}}\\
a_\sigma &=& S(S+1)-m_\sigma(m_\sigma+1)\\
m_\sigma &=& z_\sigma\langle S^z\rangle\qquad z_\sigma=\delta_{\sigma\uparrow}-\delta_{\sigma_\downarrow}
\end{eqnarray}
The ISA is an approximation for the case of low band occupation. Since we use it for the description of an insulator conduction band this is no problem, of course.\\
Due to the complicated structure of the self-energy, especially the Green's function in the denominator, one has to be careful when applying the analytic continuation rules. In appendix \ref{sec:appG} a method is shown how it can be incorporated in the Keldysh formalism.

\section{Metal/Insulator/Metal-System}\label{Sec:MIM}
In each of the three regions of the tunnel junction there is only one s-like band which is described in a tight-binding picture. Therefore each band is completely determined by its band center, band width and the lattice type. We only consider simple cubic lattices in this paper. The magnetization will be used as a model parameter, i.e. it is not calculated self-consistently. It can be approximated by a Brillouin function\cite{NoltingMag1}:
\begin{eqnarray}
\langle S_z\rangle &=& \frac{2S+1}{2S\tanh{\left((2S+1)\frac{x}{2S}\right)}}-\frac{1}{2S\tanh{\left(\frac{x}{2S}\right)}},\\
x &=& \frac{3S}{S+1}\frac{T_C}{T}\langle S_z\rangle
\end{eqnarray}
The change of total spin $S$ will qualitatively be the same as a change of temperature. Therefore we assume the spin to be constant: $S=\frac{7}{2}$. Furthermore we choose the zero point of energy to coincide with the band centers of the leads when no voltage is applied: $BC_{met}\equiv 0$. Then there are seven remaining model parameters: the band widths of the metals $W_{met}$ and the insulator $W_{ins}$, the band center of the insulator $BC_{ins}$, the band occupation in the metals $n$, the exchange coupling $J$, the tunnel coupling $\epsilon_{\mathbf{k}_M\mathbf{k}_I}^n$ and the number of insulator layers $d$. The last two parameters play a special role because for the description of a concrete experiment all the other ones are fixed. The number of layers can be determined approximately by the thickness of the sample and the lattice constant. The tunnel coupling is the only really free parameter in the model and can be used to fit theoretical to experimental results. We assume that it is independent of the wave vectors $\epsilon_{\mathbf{k}_M\mathbf{k}_I}^n\equiv \epsilon_M^n$ and choose the following ansatz for the tunnel coupling:
\begin{eqnarray}
\epsilon_L^n &=& \epsilon \lambda^d e^{-\lambda(n-1)}\quad(n\in 1,\dots,d)\\
\epsilon_R^n &=& \epsilon \lambda^d e^{-\lambda(d-n)}
\end{eqnarray}
The exponential term is responsible for damping the coupling between the metals and insulator layers that are further away. $\lambda^d$ will decrease the overall strength of the coupling with increasing thickness of the insulator for $\lambda<1$. The disadvantage of this ansatz is that one gets another undetermined parameter into the theory, namely $\lambda$. One way to determine it is by comparison with other theories. Simmons\cite{Simmons} derived a tunneling current formula through a thin insulating film. For intermediate voltages ($0<V<\phi_0$, where $\phi_0$ is the average barrier height) it reads
\begin{eqnarray}
J(d) &=& \frac{A}{d^2} \Bigl((\phi_0 - \frac{eV}{2}) e^{-Bd\sqrt{\phi_0 - \frac{eV}{2}}}-\nonumber\\*
&-&(\phi_0 + \frac{eV}{2}) e^{-Bd\sqrt{\phi_0 + \frac{eV}{2}} }\Bigr)
\end{eqnarray}
A and B are constants. By comparing the results of our theory with this formula, we find an almost perfect fit for $\lambda=0.6$. The dependence of the tunneling current on the remaining constant $\epsilon$ is generally complicated, but it was found numerically that the current scales with $\epsilon^4$ for small values of $\epsilon<0.3$ eV. Therefore it can formally be included in the unit of current together with the number of lattice sites in the insulator $N_i$:
\begin{equation}
[J] = \frac{e\epsilon^4N_i}{\hbar}
\end{equation}
\begin{figure}
\includegraphics[width=\linewidth,keepaspectratio]{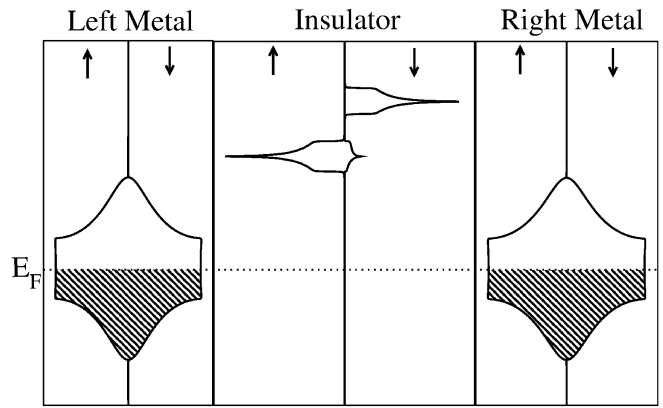}
\caption{QDOS of a M/I/M system without tunnel coupling for $T=0$.}\label{fig:princMIM1}
\end{figure}
\begin{figure}
\includegraphics[width=\linewidth,keepaspectratio]{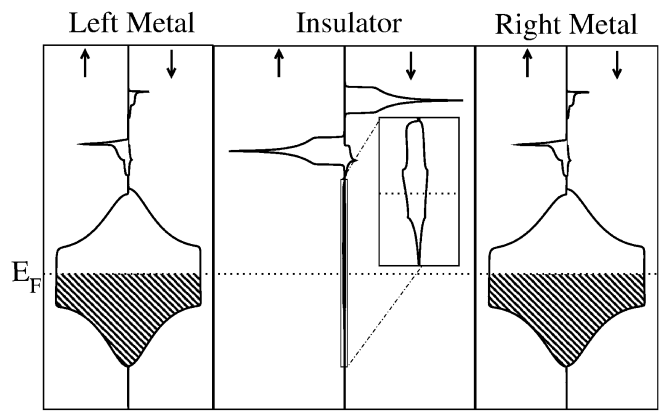}
\caption{QDOS of a M/I/M-Systems with finite tunnel coupling for $T=0$. The insulator states near the Fermi energy are additionally enlarged for clarity.}\label{fig:princMIM2}
\end{figure}
In this section both leads consist of the same normal metal. Before beginning a systematic evaluation of the model it is useful to get a better understanding of how hybridization can be used to model tunneling. In Figs. \ref{fig:princMIM1} and \ref{fig:princMIM2} the QDOS of the three regions are shown, with and without tunnel coupling $\epsilon$. Since there is no direct contact between the two leads, electrons can only get from the left to the right metal by hopping through the insulator. In the case $\epsilon=0$ this is not possible, because there are no allowed states in the energetic region of the occupied metal states. If the tunnel coupling is turned on, there will be allowed metal states in the energetic region of the insulator and vice versa. Therefore hopping between the layers becomes possible. Of course there still won't be a net current without an applied voltage.

\subsection{One-layer insulator}
\begin{figure}
\includegraphics[width=.8\linewidth]{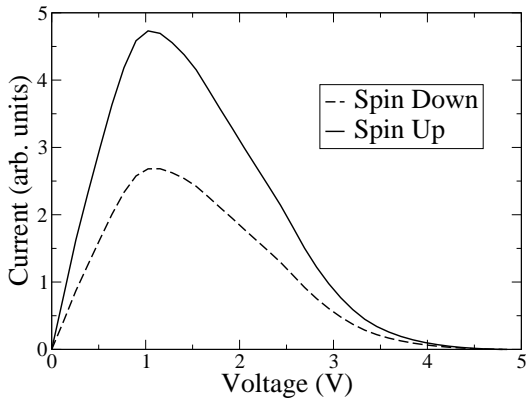}
\caption{Qualitative behavior of the spin-dependent tunneling current for a one-layer insulator. Parameters: $W_{met}=5 eV, W_{ins}=1 eV, BC_{ins}=3 eV, n=0.5, \frac{T}{T_C}=0, J=0.3 eV$}\label{fig:typcurr}
\end{figure}

\begin{figure}
\includegraphics[width=.9\linewidth]{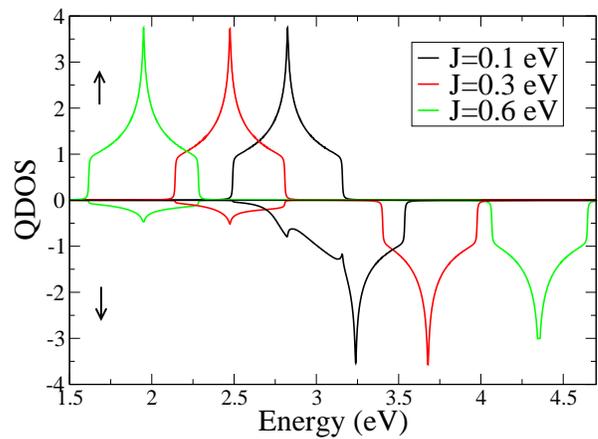}
\caption{(Color Online) Insulator QDOS for different values of the exchange coupling constant $J$ without coupling to the metals. Parameters: $W_{ins}=1 eV, \mbox{BC}_{ins}=3 eV, S=\frac{7}{2}, T=0 K$\\}\label{fig:ISA_J}
\end{figure}

\begin{figure}
\includegraphics[width=.9\linewidth]{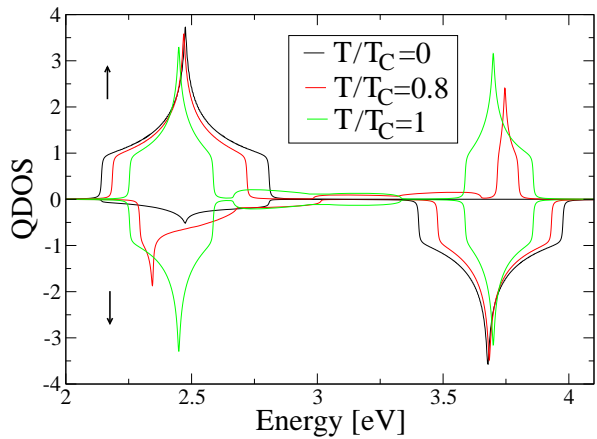}
\caption{(Color Online) Insulator QDOS for different values of temperature $\frac{T}{T_C}$ without coupling to the metals. Parameters: $W_{ins}=1 eV, \mbox{BC}_{ins}=3 eV, S=\frac{7}{2}, J=0.3 eV$}\label{fig:ISA_T}
\end{figure}

\begin{figure*}
\includegraphics[width=.9\linewidth]{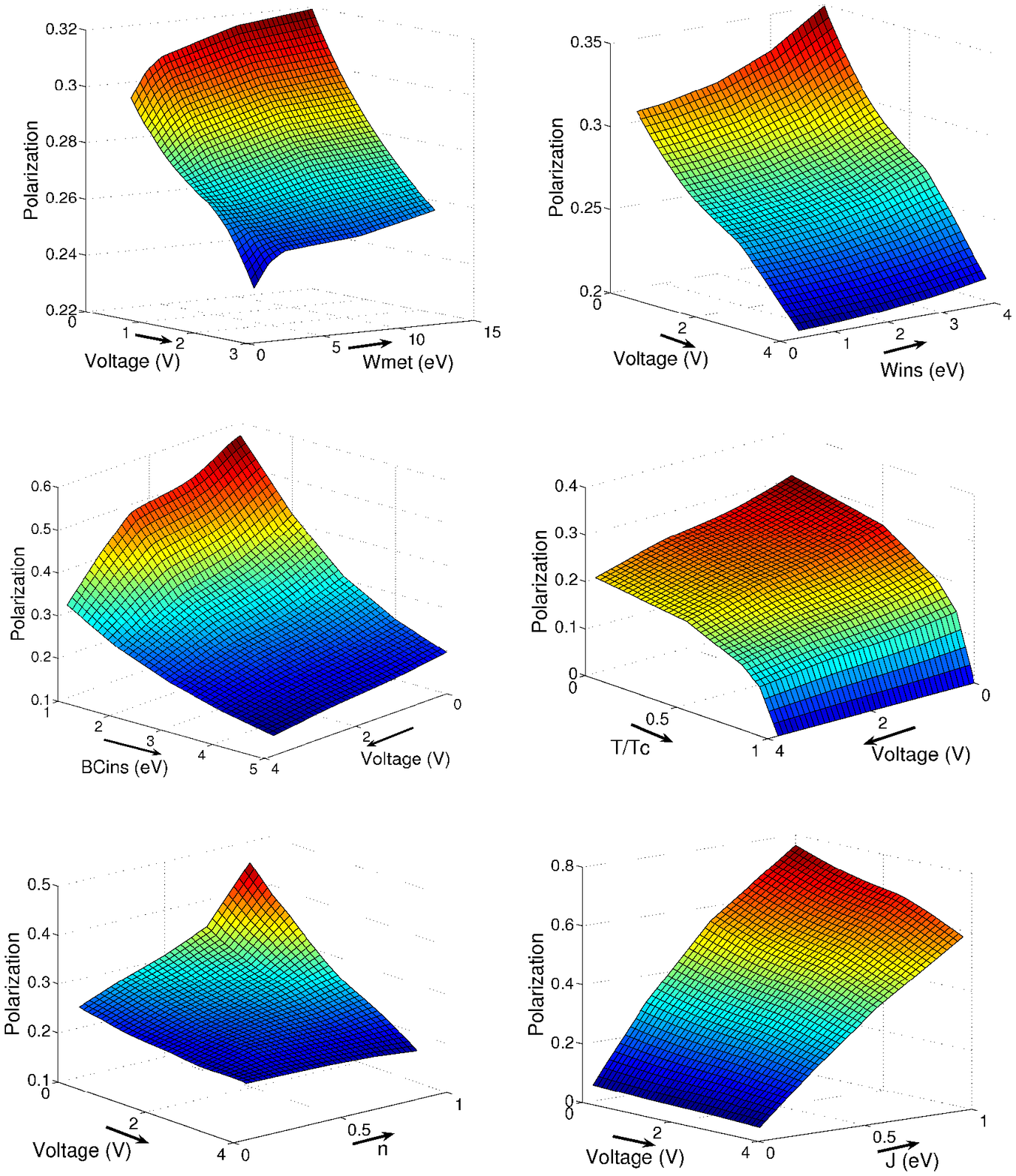}
\caption{(Color Online) Spin polarization as a function of voltage and model parameters. Blue (dark gray) corresponds to low, yellow (white) to intermediate and red (light gray) to high polarization. For better clarity it is marked by arrows in which direction the scales increase. Parameters: $W_{met}=5 eV, W_{ins}=1 eV, BC_{ins}=3 eV, n=0.5, T=0 K, J=0.3 eV$}\label{fig:pol_d=1}
\end{figure*}

A typical current profile for a one-layer insulator is shown in Fig. \ref{fig:typcurr}. For small voltages the current increases in a linear way up to a maximum and later on it goes back to zero. Without an applied voltage there are only occupied or unoccupied states lying opposite to each other. Due to the Pauli principle there can't be any tunneling in this case. By increasing the voltage the right metal gets shifted downwards compared to the left one. For small voltages the number of occupied states on the left lying opposite of unoccupied states on the right grows more or less linearly. Therefore the current increases in a linear way, too. When there is a maximal number of occupied states lying opposite to unoccupied states the current reaches its maximum. For higher voltages it decreases again and will become zero when the voltage exceeds the band width of the metal (5 eV in this case), because then there is no overlap between the two bands anymore. This bevavior is the same for both spin directions, but obviously the spin-up current is higher over the whole voltage range, i.e. there is a finite spin polarization. The reason behind this is the energetic distance between the metal and insulator bands. For the spin-up bands this distance is less than for spin-down, resulting in a stronger effect of the hybridization, which in turn will lead to a higher current.\\
\begin{figure*}
\includegraphics[width=.9\linewidth]{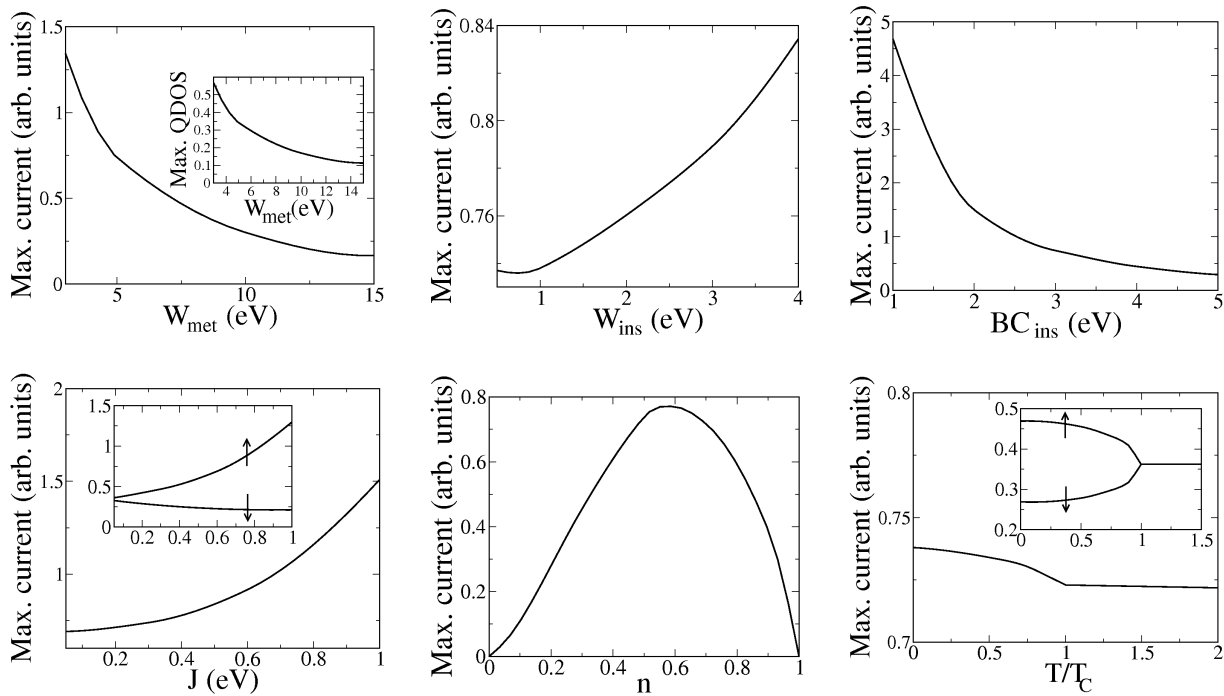}
\caption{Dependence of the tunneling current maximum on the model parameters. The inset of the $W_{met}$-figure shows the maximum of the metallic QDOS in dependence on the band width. The insets of the $\frac{T}{T_C}$- and $J$-figures show the contributions from the two spin directions. For all the other parameters they behave like the total current and are not shown for clarity. The following standard parameters were used: $W_{met}=5 eV, W_{ins}=1 eV, BC_{ins}=3 eV, n=0.5, T=0 K, J=0.3 eV$. For each picture just one of them was varied.}\label{fig:depmodpar}
\end{figure*}
In Fig. \ref{fig:depmodpar} the dependence of the maximum of the total current $J=J^\uparrow+J^\downarrow$ on the model parameters is shown, again for a one-layer system. In the case of the exchange coupling and the temperature the insets also show the spin-dependent behavior of the maximum. For all the other parameters it is qualitatively the same as the shown curves. In the following list we will explain these curves:\\
\indent Band width of the metals $W_{met}$:\\
Increasing the width of the metal bands automatically decreases the maximum of the QDOS. This is shown in the inset. A decrease of QDOS means that there are less electrons which can tunnel and at the same time less empty states which they can tunnel into. Both effects reduce the current.\\
\indent Band width of the insulator $W_{ins}$:\\
By increasing the band width of the insulator and leaving its center of gravity constant the height of the tunnel barrier gets smaller. In the model this corresponds to a higher hybridization strength because the energetic distance between metal and insulator bands decreases. So one expects an increase of current.\\
\indent Band center of the insulator $BC_{ins}$:\\
The explanation of this effect is similar to the band width. The band center of the insulator is directly proportional to the height of the tunnel barrier. Shifting the band to higher energies will lower the current.\\
\indent Exchange coupling $J$:\\
The dependence of the current on $J$ can only be understood by looking at the spin-dependent contributions shown in the inset. The spin-down current decreases slightly while the spin-up current shows a strong increase resulting in an increase of the total current. This behavior can be explained with the help of the $BC_{ins}$-curve. For $J=0$ eV both spin-down and spin-up bands have the same center of gravity. Increasing $J$ shifts the spin-up band to lower and the spin-down band to higher energies (compare Figs. \ref{fig:expl_J} and \ref{fig:ISA_J}). It can immediately be seen from the picture that there will be a slight decrease in spin-down current and a bigger increase of spin-up current.\\
\indent Band occupation $n$:\\
If the metal band is completely occupied (n=1 for both spin directions) or completely empty, it becomes an insulator and the current vanishes. For small $n$ there is only a small amount of electrons in the left metal which can tunnel. Vice versa, for high $n$ there is only a small amount of unoccupied states in the right metal which can be tunneled into. For half filling there is an unoccupied state in the right metal for every electron in the left metal. Therefore one would expect a symmetric curve, with positive slope for $n<0.5$, a maximum at $n=0.5$ and negative slope for $n>0.5$. But the result of the model calculation shows that the curve is shifted asymmetrically to higher band occupations. The reason behind this is that the distance between Fermi energy and the lower band edge of the insulator is smaller for high $n$. Therefore the symmetry between small and high $n$ is broken.\\
\indent Temperature $T/T_C$:\\
With increasing temperature spectral weight is shifted from the lower to the upper spin-up band (compare Fig. \ref{fig:ISA_T}). This effectivly increases the tunnel barrier for spin-up electrons and thus the current is decreased. With a similar argument it can be understood why the spin-down current will increase with increasing temperature. At the Curie temperature $T_C$ both curves meet and above $T_C$ the current is practically no longer dependent on temperature.\\
\begin{figure}
\includegraphics[width=.9\linewidth]{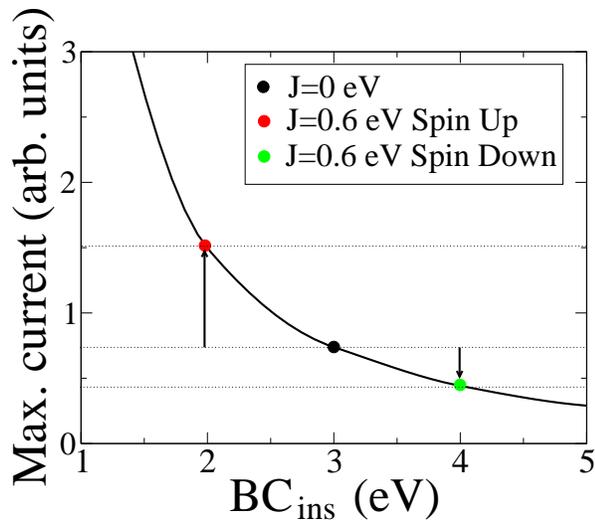}
\caption{(Color Online) Graphical explanation of the behavior of the current maximum in dependence on the exchange coupling $J$ with the $BC_{ins}$-curve from Fig. \ref{fig:depmodpar}}\label{fig:expl_J}
\end{figure}
The explanation of the $J$-curve with Fig. \ref{fig:expl_J} will be crucial for the understanding of the spin polarization $P=\frac{J^\uparrow-J^\downarrow}{J^\uparrow+J^\downarrow}$. Since the slope of the $BC_{ins}$-curve decreases drastically for higher band centers, it follows that higher band centers (or tunnel barriers) are bad from the point of view of the spin polarization.\\
In Fig. \ref{fig:pol_d=1} the spin polarizations are shown as functions of the applied voltage and the model parameters. The spin polarization decreases with increasing voltage for all parameters. In the case of one insulating layer only the band center of the right metal is influenced by the voltage, while the band centers of the insulating layer and the left metal remain constant. The voltage shifts the right band center to lower energies, i.e. the height of the tunnel barrier on the right side increases. With the argument given in the paragraph above this leads to a lower spin polarization. The dependence on the model parameters will be discussed in the following list:\\
\indent Band width of the metals $W_{met}$:\\
The spin polarization only shows a weak dependence on $W_{met}$. It changes only about 2 \% during a change of band width from 3 to 15 eV. The apparent sudden tilt for high voltages and low band widths results from numerical difficulties since the current is almost zero for those parameters.\\
\indent Band width of the insulator $W_{ins}$:\\
High insulator band widths are good for spin polarization. This is especially obvious for small voltages. For higher voltages this increase is not very pronounced.\\
\indent Band center of the insulator $BC_{ins}$:\\
The spin polarization changes drastically in dependence on $BC_{ins}$, especially for the case of low band widths (notice the spin polarization scale!). For low voltages it grows by a factor of three from about 20 \% to almost 60\% when decreasing the band center from 5 to 1 eV.\\
\indent Temperature $T/T_C$:\\
The spin polarization closely resembles the behavior of the magnetization, i.e. it has a saturation value for $T=0$, slowly decreases with increasing $T$ and falls off very rapidly near the Curie temperature. Above $T_C$ it is zero, as expected.\\
\indent Band occupation $n$:\\
The spin polarization is constant for low $n$. For very high $n$ and low voltages there is a clear increase since the tunnel barrier is lowered with increasing $n$. This is good for high polarization as explained above.\\
\indent Exchange coupling $J$:\\
The spin polarization shows the strongest change in dependence on $J$. By increasing the exchange coupling one can reach polarizations of almost 80 \% (although such a high $J$ might not be realistic with the chosen band width of 1 eV). The increase is approximately linear in $J$.

\subsection{Two-layer insulator}
\begin{figure}
\includegraphics[width=\linewidth]{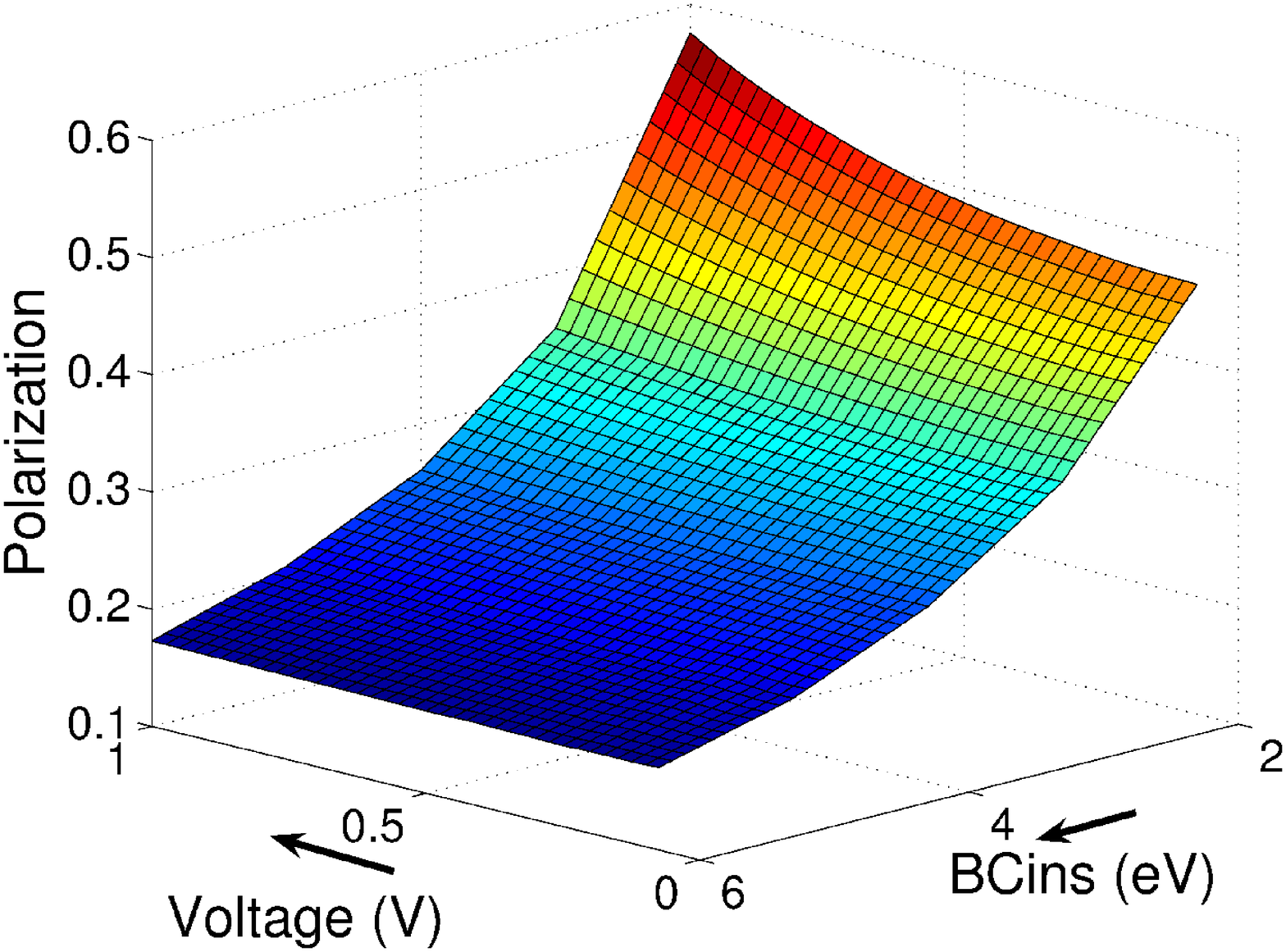}
\caption{(Color Online) Spin polarization for a two-layered insulator as a function of voltage and the band center of the insulator $BC_{ins}$. The arrows mark in which direction the scales increase. Parameters: $W_{met}=5 eV, W_{ins}=1 eV, n=0.5, T=0 K, J=0.3 eV$}\label{fig:pol_d=2}
\end{figure}
If we replace the one-layer insulator by a two-layer insulator the dependence of the spin polarization on the model parameters remains the same but the dependence on the voltage is reversed. This is shown exemplary in Fig. \ref{fig:pol_d=2} for $BC_{ins}$. Now the spin polarization \emph{increases} with increasing voltage. If we denote the metals by $L$ and $R$ as above and the left insulator layer by $I1$ and the right layer by $I2$ the following happens when the voltage is increased: the distance between $L$ and $I1$ and $R$ and $I2$ stays the same. At the same time the distance between $L$ and $I2$ is decreased and the distance between $R$ and $I1$ is increased. The first effect will increase the spin polarization (lower barrier height!) while the second one will decrease it. Due to the high slope of the $BC_{ins}$-curve for low band centers the first effect dominates and leads to the observed behavior of the spin polarization.\\
Insulators with more than two layers behave qualitatively in the same way as discussed here. Only the strength of the current falls off with increasing number of layers.\\
DeWeert and Girvin\cite{DeWeert} performed a Boltzmann-equation study of M/I/M-tunneling where they also found that spin polarization will increase with voltage. This is not surprising since they made assumptions similar to ours. However, they treated the ferromagnetism in a phenomenological way by introducing different tunneling life times for the two spin directions. In our case the magnetism is a direct consequence of the considered microscopic model.

\subsection{Comparison to experiment}
\begin{figure}
\includegraphics[width=.8\linewidth]{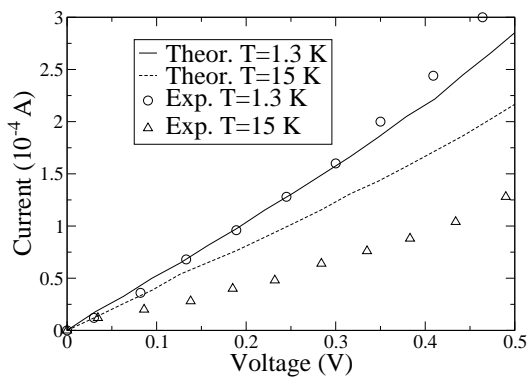}
\caption{Comparison of the theoretical results (lines) with experimental measurements by Hao et al.\cite{Moodera90} (circles and triangles). Parameter for the calculation: $W_{met}=26 eV, W_{ins}=0.9 eV, BC_{ins}=-1.3 eV, n=0.33, J=0.1 eV, d=3$}\label{fig:MIM_Moodera}
\end{figure}
Now we want to compare our theoretical results with experimental measurements for an Al/EuS/Al-system\cite{Moodera90}. In Fig. \ref{fig:MIM_Moodera} the results of the experiment (circles and triangles) are shown together with our calculation (lines). The model parameters are taken from Wachter\cite{Wachter}. The lattice constant of bulk EuS is $a_0=5.97$ \r{A}. In the experiment the insulator had a thickness between 17.6 \r{A} and 21.9 \r{A} which amounts to approximately three layers. Of course this is an approximation since the lattice constant will change for small thicknesses, but the order of magnitude should be the same. The parameters for Al were taken from Papaconstantopoulos\cite{Bandstructure}. The tunnel coupling $\epsilon$ is the last unknown parameter. The unit of current is $\frac{e}{\hbar}\epsilon^4 N_i$, therefore $\epsilon$ can be used to fit the intensity of the theoretical current to the experimental results. This was already done in Fig. \ref{fig:MIM_Moodera} with $\epsilon^4 N_i = 5.3\cdot 10^6$. The surface area of the sample was $A=3\cdot10^{-3}cm^2$. Assuming that the lattice sites are ordered quadratically one can estimate the number of lattice sites to $N_i=8.4\cdot 10^{11}$. Hence $\epsilon$ is known: $\epsilon\approx 0.05$ eV. Of course, this is a strongly simplified approximation, but it shows that the tunnel coupling is small enough to justify its inclusion in the unit of current, which is only possible for $\epsilon<0.3$ eV.\\
There are several possible reasons for the quantitative difference between theory and experiment. Aluminum is not well described by a single band tight-binding approximation. Since the metal QDOS has a strong influence on the current characteristics one has to expect differences. Another reason is the treatment of the tunnel barrier. It was assumed to be rectangular and was rigidly tilted by the applied voltage. This assumption is, especially for reasons of simplicity, often made. But on the other hand it is known to be a very crude approximation for real barriers since they can have varying barrier heights and even holes\cite{Gross}. Another quantitative difference is the behavior in dependence on temperature. Both in experiment and theory the slope of the curves decreases with increasing temperature, but this decrease is stronger for the experimental curves. The origin of this difference most likely lies in the treatment of the tunnel coupling. It was already shown that spin-up and spin-down current approach each other with increasing temperature, but that the spin-up current decreases a little bit faster than the spin-down current increases. Thus the total current decreases, but not enough to explain the experimental results. This decrease would be larger if the $BC_{ins}$ curve in Fig. \ref{fig:depmodpar} would be steeper. That might be accomplished by explicitly taking into account the wave-vector dependence of the tunnel coupling which was neglected in this paper.

\section{Metal/Insulator/Superconductor-System}\label{Sec:MIS}
\begin{figure}
\includegraphics[width=\linewidth]{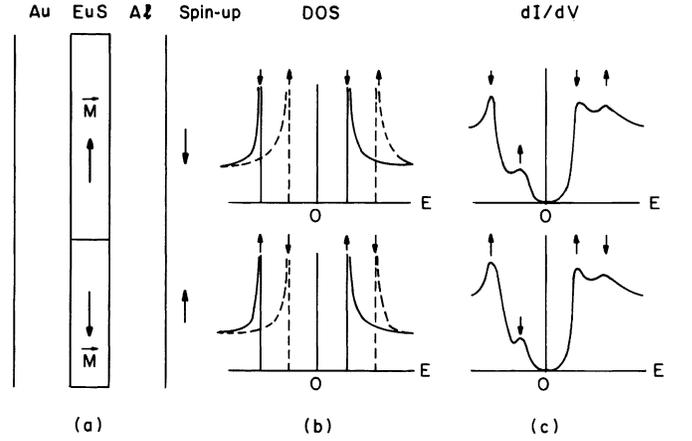}
\caption{Contributions to the conductance from two different domains of the EuS (from Hao et al.\cite{Moodera90}). (a) shows the schematic design of the Al/EuS/Au system, (b) the Zeeman-split density of states of the Al and (c) the expected conductance. Its overall behavior is the same for both domains, only the spin directions are exchanged.}\label{fig:MIS_Mood_princ}
\end{figure}

In this section we replace one of the metals with a superconductor (M/I/M $\rightarrow$ M/I/S). For the derivation of the tunneling current formula we assumed that the leads are non-interacting. Obviously, this assumption does not hold in the case treated here. $H_L$ has to be replaced by the BCS-Hamiltonian\cite{Zagoskin}:
\begin{eqnarray*}
H_L \rightarrow H_L &=& \sum_{\substack{M\\M\in L,R}}\sum_{\mathbf{k}_M\sigma} \epsilon_{\mathbf{k}_M} c_{\mathbf{k}_M\sigma}^+ c_{\mathbf{k}_M\sigma} -\\*
&-& \Delta\sum_{\mathbf{k}_R}(c_{-\mathbf{k}_R\downarrow}c_{\mathbf{k}_R\uparrow} + c_{\mathbf{k}_R\uparrow}^+c_{-\mathbf{k}_R\downarrow}^+) + \frac{\Delta^2}{V}
\end{eqnarray*}
$\Delta$ is the bandgap of the superconductor. The problem with this approach lies in the fact that $H_L$ does not commute with the particle number operator anymore. In principle one has to derive a new current formula with this new $H_L$ as was done e.g. by Zeng et al.\cite{Zeng}. Here we use an alternative approach, which is known as the effective medium approach. The basic idea is to split off the interacting part and use it for a renormalization of the one-particle energies and therefore for the density of states:
\begin{equation}
\rho^{(0)}(E) \rightarrow \rho^{(S)}(E)
\end{equation}
$\rho^{(S)}(E)$ is the superconducting QDOS. Thus the model Hamiltonian remains the same, only the QDOS has to be changed in the numerical treatment. The disadvantage of this approach is the restriction to single electron effects, i.e. the neglicence of Cooper pair tunneling for example, but the comparison with experiment will justify this.\\
In recent years a lot of research was focused on the transport properties of ferromagnet-superconductor junctions. Especially the role of Andreev reflection has been investigated by many authors\cite{deJong}\cite{Soulen}\cite{Shi}\cite{Zutic}\cite{Buzdin}. Since Andreev reflection involves more than one electron it cannot be modeled by our effective medium approach. Again this can be justified by comparison with experiment later in this section.\\
The main difference between normal metals and superconductors is the appearance of the bandgap. Its width is usually of the order of meV. Therefore we can restrict the discussion to voltages which correspond to these energies. For higher voltages the results will coincide with the M/I/M curves. The reason for introducing the multi-layer insulator was the modelling of the band tilting by the applied voltage. In the case considered here this should have no effect on the current, since the tilt is of the order of meV with a total barrier height in the order of eV. The change from a one-layer to a many-layer system will merely change the intensity but not the qualitative behavior of the current.\\
\begin{figure}
\includegraphics[width=\linewidth]{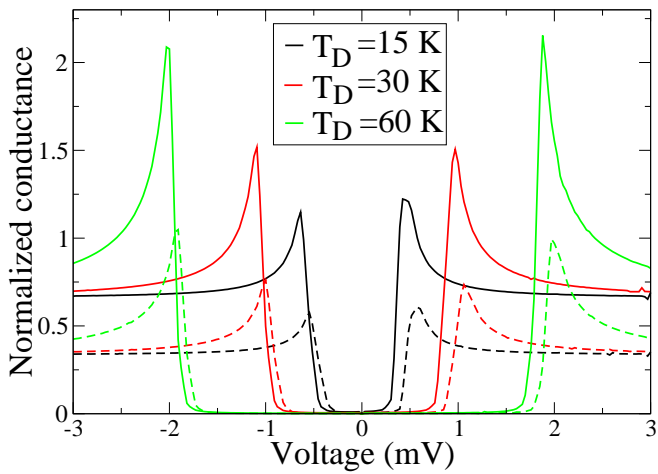}
\caption{(Color Online) Differential conductance of the M/I/S system for different values of the Debye temperature. Dotted lines are spin-down, continuous lines are spin-up conductances. All curves are normalized according to Eq. (\ref{equ:norm}). Parameters: $W_{\mbox{met}}=3 eV, W_{\mbox{ins}}=1 eV, \mbox{BC}_{\mbox{ins}}=3 eV, \epsilon= 0.03 eV, T=0 K, n=0.5, J=0.3 eV$}\label{fig:MIS_cond_TD}
\end{figure}
For better comparison with experiment later on we focussed on the differential conductance $\frac{dJ^\sigma}{dV}$ rather than the current itself. It is normalized on the total conductance of the M/I/M system
\begin{equation}\label{equ:norm}
\frac{\frac{dJ^\sigma}{dV_{MIS}}}{\frac{dJ^\uparrow+dJ^\downarrow}{dV_{MIM}}}.
\end{equation}
In Fig. \ref{fig:MIS_cond_TD} the conductance is shown for several values of the Debye temperature $T_D$. For the parameters used we get the following bandgaps:
\begin{eqnarray}
\Delta(T_D=15 K)=0,46 \mbox{ meV}\nonumber\\
\Delta(T_D=30 K)=0,92 \mbox{ meV}\\
\Delta(T_D=60 K)=1,85 \mbox{ meV}\nonumber
\end{eqnarray}
There is no current for voltages lower than these bandgaps. Furthermore the conductance curves are not symmetric about the origin: for positive voltages the spin-up current starts first followed by the spin-down current, for negative voltages vice versa. Ignoring this spin splitting the basic form of the conductance can be understood with the following arguments: for $V=0$ the chemical potentials on both sides are equal and current cannot flow. With increasing voltage there still won't be any current since there are no states available in the superconductor. Only when the voltage exceeds the bandgap $|eV|>\Delta$ there will be occupied and unoccupied states opposite to each other. Hence one expects an increase of conductance at the edge of the bandgap. Since there are many available states above it this increase will be stronger than in the M/I/M system.\\
\begin{figure}
\includegraphics[width=0.8\linewidth]{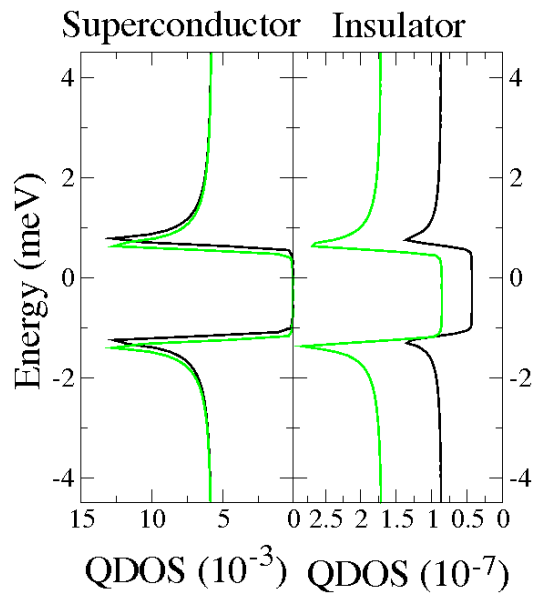}
\caption{(Color Online) QDOS of the superconductor and the insulator in the range of the bandgap. There is a clear splitting between spin-up (green/light gray) and spin-down (black) observable. Parameters as in Fig. \ref{fig:MIS_cond_TD}\\}\label{fig:MIS_qdos}
\end{figure}
The spin splitting of the conductance curves is due to the coupling between the ferromagnetic insulator and the non-magnetic superconductor. The hybridization repels the two bands. The spin-up insulator band is closer to the superconductor band and therefore causes a larger band shift. This can be seen from the QDOS in Fig. \ref{fig:MIS_qdos}. For positive energies the band edge of the spin-up band is just below the bandgap while the spin-down band is just above it. When applying a positive voltage one expects the spin-up current to start first which was indeed observed in Fig. \ref{fig:MIS_cond_TD}. The QDOS of the insulator is finite inside the bandgap, i.e. there will be a finite contribution to the tunneling current from this region! This effect has its origin in the hybridization with the (constant) QDOS of the metal and is physically not explainable. Though the current inside the bandgap is finite, it is still at least three orders of magnitude lower than the current above. Thus it cannot be seen in Fig. \ref{fig:MIS_cond_TD}.\\
\begin{figure}
\includegraphics[width=.7\linewidth]{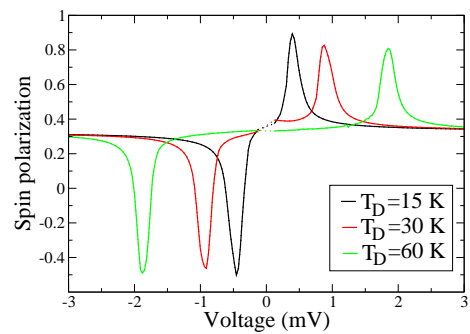}
\caption{(Color Online) Spin polarization of the M/I/S junction for different values of the Debye temperature. The dotted region around the origin was omitted since the spin polarization is not defined there. The same parameters as in Fig. \ref{fig:MIS_cond_TD} were used.\\}\label{fig:MIS_pol_TD}
\end{figure}
Compared to the M/I/M system the spin polarization shows a richer structure (Fig. \ref{fig:MIS_pol_TD}). Due to the different onsets of the current for the two spin directions there are two strong peaks at $\pm\Delta$. For positive voltages one observes positive spin polarization with a maximum of about 90 \%. It decreases slightly with increasing width of the bandgap. For negative voltages the spin polarization will be negative in a narrow region with a minimum of about -50 \%.\\
\begin{figure}
\includegraphics[width=.7\linewidth]{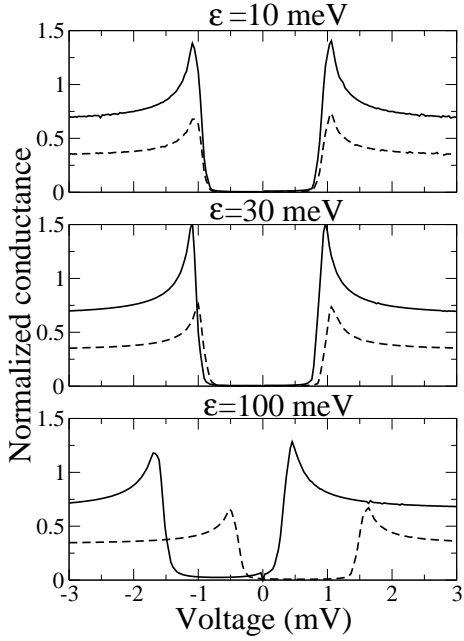}
\caption{Differential conductance of the M/I/S system for different values of the tunnel coupling $\epsilon$. Dotted lines represent spin-down, continuous lines represent spin-up conductances. They are normalized according to Eq. (\ref{equ:norm}). Parameters: $W_{\mbox{met}}=3 eV, W_{\mbox{ins}}=1 eV, \mbox{BC}_{\mbox{ins}}=3 eV, T_D= 30 K, T=0 K, n=0.5, J=0.3 eV$}\label{fig:MIS_cond_eps}
\end{figure}
\begin{figure}
\includegraphics[width=.7\linewidth]{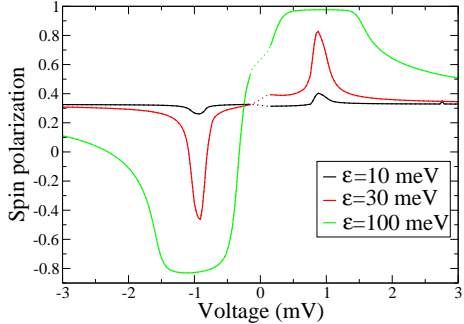}
\caption{(Color Online) Spin polarization of the M/I/S system for different values of the tunnel coupling $\epsilon$. The spin polarization is not well defined in the dotted region around the origin. The same parameters as in Fig. \ref{fig:MIS_cond_eps} were used.\\}\label{fig:MIS_pol_eps}
\end{figure}
The tunnel coupling $\epsilon$ determines the strength of the hybridization between the leads and the insulator. The bigger it gets, the stronger the spin splitting of the superconductor should be. This effect is confirmed in Fig. \ref{fig:MIS_cond_eps}. For $\epsilon=10$ meV there is almost no spin splitting observable and the spin polarization only has two small bumps accordingly. By increasing $\epsilon$ the spin splitting is increased until the bandgaps of spin-up and spin-down almost do not coincide anymore for $\epsilon=100$ meV. Thus in this case the highest spin polarization of about 95 \% is reached which is also stable across a broad voltage range of about 1 meV.\\
\begin{figure}
\includegraphics[width=.7\linewidth]{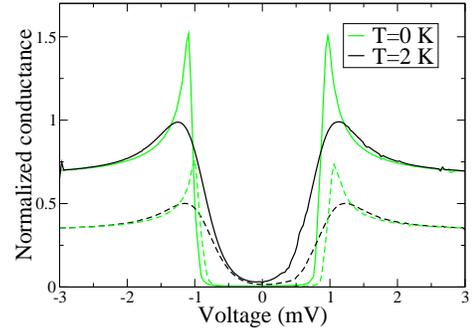}
\caption{(Color Online) Differential conductance of the M/I/S system as a function of the applied voltage for two different temperatures. Continuous lines represent spin-up, broken lines spin-down conductances. Parameters: $W_{\mbox{met}}=3 eV, W_{\mbox{ins}}=1 eV, \mbox{BC}_{\mbox{ins}}=3 eV, T_D= 30 K, n=0.5, J=0.3 eV$}\label{fig:MIS_T}
\end{figure}
\begin{figure}
\includegraphics[width=.7\linewidth]{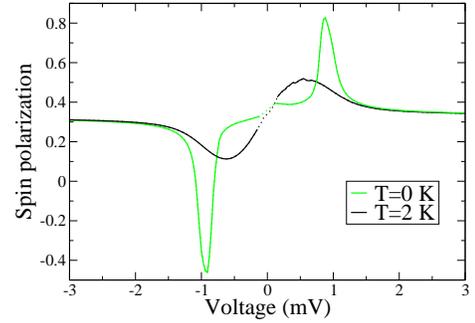}
\caption{(Color Online) Spin polarization of the M/I/S junction for two different temperatures. The spin polarization is not well defined in the dotted region around the origin. Parameters as in the figure above.}\label{fig:MIS_pol_T}
\end{figure}
The last model parameter which will be discussed is the temperature $T$. In Figs. \ref{fig:MIS_T} and \ref{fig:MIS_pol_T} two conductance curves and the according spin polarizations are shown for $T=0$ and $T>0$. Increasing the temperature has two effects: first the bandgap is reduced which is equivalent to a reduction of $T_D$ and thus nothing new. Second the Fermi edge is softened and this can lead to finite occupation above the bandgap. Therefore current can already flow for voltages below the bandgap which leads to the observed softening of the conductance curves. For the same reason the maximum of the conductance is lower.\\
\begin{figure}
\includegraphics[width=\linewidth]{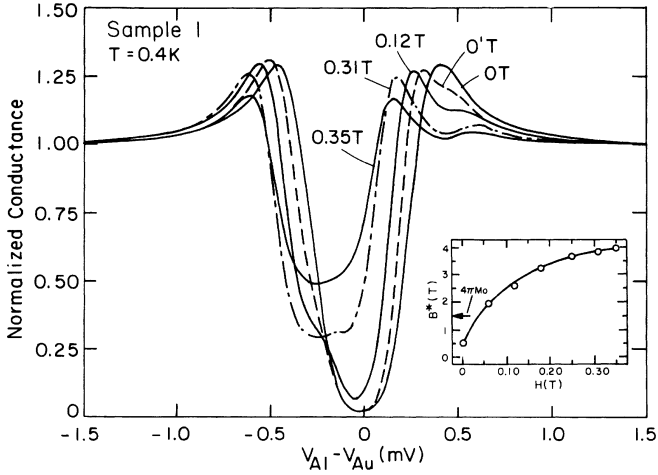}
\caption{Experimental conductance curves for an Al/EuS/Au tunnel junction and different strengths of the applied magnetic field (from Hao et al.\cite{Moodera90}). The inset shows the dependence of the internal magnetic field $B^*$ on the applied field $H$.}\label{fig:MIS_Mood_exp}
\vspace{0.5cm}
\end{figure}
The results of the model calculation will now be compared to experimental results obtained in the group of Moodera\cite{Moodera90}\cite{Moodera93}\cite{Moodera04}. They used several europium chalkogenides as tunnel barriers. All the results are qualitativly the same, thus we will compare our results with the EuS measurements\cite{Moodera90}  only. The tunnel junction consisted of an Au/EuS/Al system where Al becomes superconducting below $T=1.2$ K. In Fig. \ref{fig:MIS_Mood_exp} the experimentally measured conductance is shown for several values of the applied magnetic field. Even without applied field they measured a finite Zeeman splitting which corresponds to an internal field of $B^*=0.5$ T. The dependence of this internal field on the applied field is shown in the inset of the figure. Obviously the internal field is much bigger than can be explained by the applied field alone. This is a strong indication of the coupling between EuS and the Al-lead.\\
\begin{figure}
\includegraphics[width=\linewidth]{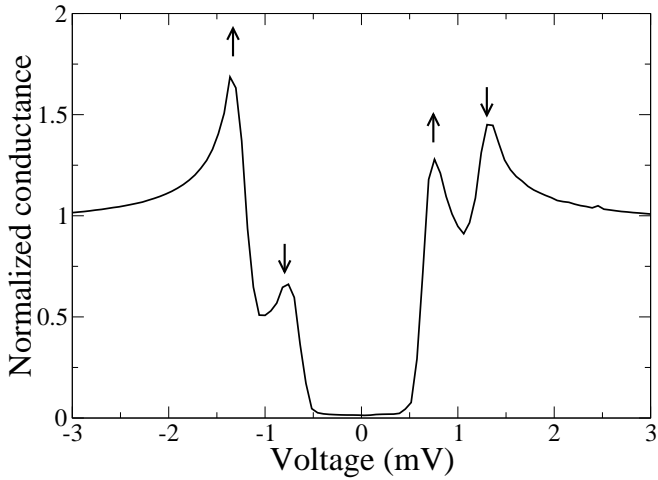}
\caption{Normalized total conductance as a function of the applied voltage. The arrows mark the spin direction which contributes most to the respective maximum. The width of the bandgap is about 0.9 meV. Parameters: $W_{met}=3 eV, W_{ins}=1 eV, BC_{ins}=3 eV, n=0.5, T=0, \epsilon=70 meV, J=0.3 eV, T_D=30 K$}\label{fig:MIS_compexp}
\end{figure}
In our model the internal field is created by the hybridization between the ferromagnetic insulator and the paramagnetic leads. Therefore an increase of the tunnel coupling in the model corresponds to applying an external field in the experiment. In Fig. \ref{fig:MIS_compexp} the theoretical total conductance is shown for $\epsilon=70$ meV. The model calculation for the smaller $\epsilon=30$ meV in Fig. \ref{fig:MIS_T} agrees qualitatively with the $H=0$ T-curve. Furthermore there is also an agreement between the curves with applied external field and the results for $\epsilon=70$ meV. The authors also discuss their results in terms of the superconductor density of states according to Fig. \ref{fig:MIS_Mood_princ}. Obviously there is a good qualitative agreement between their expectation and the results of our model calculation.\\
Andreev reflections would show in the conductance curve as a characteristic peak near $V=0$ (compare e.g. the conductance curves measured by Soulen et al.\cite{Soulen}). This peak was not measured by Hao et al. and it cannot appear in our model as was explained above. Thus we can draw the conclusion that Andreev reflection is strongly suppressed in this system which is a consequence of the high spin polarization. Hence only single-particle excitations contribute to the conductance and the effective medium approach should be sufficient.

\section{Summary and Outlook}\label{Sec:Conclusion}
The topic of this paper was the theoretical modelling of spin filter experiments, i.e. the spin-dependent tunneling through a ferromagnetic insulator. Due to the coupling between two outer leads and the insulator with in general different chemical potentials the tunnel junction is not in thermal equilibrium anymore. Thus one has to use the non-equilibrium extension of the usual many-body theory, the so-called Keldysh formalism. We presented a model which can be used to simulate tunneling of electrons through a potential barrier in the presence of interactions. The insulating region was described by the Kondo Lattice Model and treated within an interpolating self-energy approach. It can be modelled as a one- or multi-layered system. The leads consisted of either two normal metals or one normal metal and a superconductor. The metals were assumed to be non-interacting, while we used the BCS theory to describe the superconductor. The tunneling process itself was simulated by a hybridization between the conduction bands of the leads and the insulator. Following Haug and Jauho\cite{Haug} we derived a current formula which holds for non-interacting leads and arbitrary interactions in the insulator. In all considered systems a finite spin polarization was found. The degree of polarization was strongly dependent on the model parameters. It was possible to explain the current characteristics with means of the QDOS. The comparison with experiment has shown that the model predicts the right qualitative behavior, but that there are also clear quantitative disagreements. They stem from the not sufficiently realistic treatment of the tunnel coupling and the band structure of the used materials. It was especially suspected that the tunneling probability decreases too slowly with increasing barrier height. Therefore the spin polarization is probably underestimated.\\
The superconducting contacts were treated as an effective medium. A strong spin splitting of the superconductor QDOS was observed which was due to the coupling with the ferromagnetic insulator. Thereby obvious maxima and minima of spin polarization formed near the edges of the bandgap. Even negative polarization could be observed. Comparison with experiment showed a qualitative agreement with the model results for the M/I/S system, too.\\
There are several aspects of the model which might be improved. One is the treatment of the tunnel coupling. Especially by the comparison with experiment for the M/I/M system it became clear that the assumption of wave-vector independent tunnel coupling is too much simplified. It would be very desirable to have a method to derive the tunnel coupling from the other model parameters which are easier accessible by experiments. A second important improvement would be the use of real bandstructures, e.g. from DFT calculations.

\begin{appendix}
\section{Derivation of the Current Formula}\label{sec:appcurr}
\begin{figure}
\includegraphics[width=\linewidth]{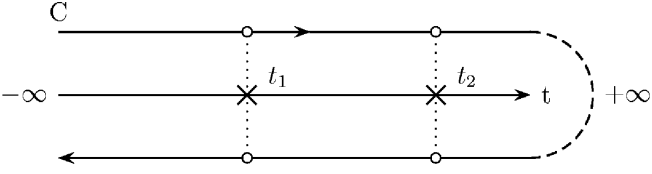}
\caption{Time contour C in the Keldysh formalism}\label{fig:contour}
\end{figure}
Evaluation of the commutator in Eq. (\ref{form:ansatzJL}) with $H$ given by Eqs. (\ref{form:ModelH})-(\ref{form:HT}) leads to
\begin{eqnarray}
J_L^\sigma(t) &=& \frac{2e}{\hbar} \sum_{\mathbf{k}_L\mathbf{k}_I n} \mbox{Re}\left(\epsilon_{\mathbf{k}_L\mathbf{k}_I}^n i\langle c_{\mathbf{k}_L\sigma}^+(t) d_{\mathbf{k}_I\sigma,n}(t)\rangle\right)\nonumber\\
&=& \frac{2e}{\hbar} \sum_{\mathbf{k}_L\mathbf{k}_I n} \mbox{Re}\left(\epsilon_{\mathbf{k}_L\mathbf{k}_I}^n G_{\mathbf{k}_I\mathbf{k}_L\sigma,n}^<(t,t)\right)\label{form:JL2}
\end{eqnarray}
Following Keldysh\cite{Keldysh} we defined the lesser Green's function in the last step:
\begin{equation}
G_{\mathbf{k}_I\mathbf{k}_L\sigma,n}^<(t,t') \equiv i\langle c_{\mathbf{k}_L\sigma}^+(t') d_{\mathbf{k}_I\sigma,n}(t)\rangle
\end{equation}
Obviously the current is known if it is possible to calculate this Green's function. This task can be accomplished by deriving the equation of motion of the corresponding causal Green's function $G_{\mathbf{k}_I\mathbf{k}_L\sigma,n}^c(t,t')=-i\langle\mathcal{T} d_{\mathbf{k}_I\sigma,n}(t) c_{\mathbf{k}_L\sigma}^+(t')\rangle$. Its formal solution is
\begin{eqnarray}
G_{\mathbf{k}_I\mathbf{k}_L\sigma,n}^c(t,t') &=& \frac{1}{\hbar} \int_{-\infty}^{\infty} dt_1 \sum_m \epsilon_{\mathbf{k}_L\mathbf{k}_I}^{m*}\cdot\nonumber\\*
&\cdot& G_{\mathbf{k}_I\sigma,nm}^c(t,t_1) g_{\mathbf{k}_L}^c(t_1,t')\label{form:Gmixed}
\end{eqnarray}
where we have used the free lead Green's function
\begin{equation}
\left(-i\hbar\frac{\partial}{\partial t'} - \epsilon_{\mathbf{k}_L}\right) g_{\mathbf{k}_L}^c(t,t') = \hbar\delta(t-t').
\end{equation}
Since the non-equilibrium Green's function (NEGF) has formally the same perturbation expansion, with integration over the real time axis changed to integration over a time contour C shown in Fig. \ref{fig:contour}\cite{Haug}, we can immediately write it down as
\begin{eqnarray}
G_{\mathbf{k}_I\mathbf{k}_L\sigma,n}(t,t') &=& \frac{1}{\hbar} \int_C d\tau \sum_m \epsilon_{\mathbf{k}_L\mathbf{k}_I}^{m*}\cdot\nonumber\\*
 &\cdot&G_{\mathbf{k}_I\sigma,nm}(t,\tau) g_{\mathbf{k}_L}(\tau,t')
\end{eqnarray}
Applying the analytic continuation rules given by Langreth\cite{Langreth} to get the lesser part of the NEGF yields
\begin{widetext}
\begin{equation}
G_{\mathbf{k}_I\mathbf{k}_L\sigma,n}^<(t,t') = \frac{1}{\hbar} \int_{-\infty}^{\infty} dt_1 \sum_m \epsilon_{\mathbf{k}_L\mathbf{k}_I}^{m*} \Bigl(G_{\mathbf{k}_I\sigma,nm}^r(t,t_1) g_{\mathbf{k}_L}^<(t_1,t') + G_{\mathbf{k}_I\sigma,nm}^<(t,t_1) g_{\mathbf{k}_L}^a(t_1,t')\Bigr)
\end{equation}
\end{widetext}
Putting this expression in (\ref{form:JL2}) and Fourier transforming the whole equation leads to the result Eq. (\ref{form:resJL}).

\section{Calculation of the Insulator Green Functions}\label{sec:appG}
The calculation of the retarded and lesser Green's function in the insulator runs along similar lines as the derivation of the current formula. The first step will be to solve the equation of motion of the corresponding causal Green's function. The Hamiltonian is again given by Eqs. (\ref{form:ModelH})-(\ref{form:HT}). This time we have to explicitly take the interactions, i.e. the KLM, into account. Its self-energy was defined by Eq. (\ref{equ:KLMself-energy}). Then the equation of motion reads
\begin{widetext}
\begin{equation}
E G_{\mathbf{k}_I\sigma,nm}^c(E) = \hbar\delta_{nm} + \sum_l\epsilon_{\mathbf{k}_I}^{nl} G_{\mathbf{k}_I\sigma,lm}^c(E) + \sum_l\Sigma_{\mathbf{k}_I\sigma,nl}^c(E) G_{\mathbf{k}_I\sigma,lm}^c(E) + \sum_{\substack{M \\ M\in L,R}}\sum_{\mathbf{k}_M} \epsilon_{\mathbf{k}_M\mathbf{k}_I}^{n*} G_{\mathbf{k}_M\mathbf{k}_I\sigma,m}^c(E)
\end{equation}
\end{widetext}
The 'mixed' Green's function $G_{\mathbf{k}_M\mathbf{k}_I\sigma,m}^c(E)$ can be calculated as in (\ref{form:Gmixed}). The final expression is very similar:
\begin{equation}
G_{\mathbf{k}_M\mathbf{k}_I\sigma,m}^c(E) = \frac{1}{\hbar}\sum_l \epsilon_{\mathbf{k}_M\mathbf{k}_I}^l g_{\mathbf{k}_L}^c(E) G_{\mathbf{k}_I\sigma,lm}^c(E)
\end{equation}
The equation of motion becomes
\begin{widetext}
\begin{equation}
\sum_l\underbrace{\left(E\delta_{nl}-\epsilon_{\mathbf{k}_I}^{nl}-\Sigma_{\mathbf{k}_I\sigma,nl}^c(E)\right)}_{\equiv \hbar\left(g_{\mathbf{k}_I\sigma}^c(E)\right)_{nl}^{-1}} G_{\mathbf{k}_I\sigma,lm}^c(E) = \hbar\delta_{nm} + \frac{1}{\hbar}\sum_{\substack{M\\M\in L,R}} \sum_{\mathbf{k}_M l} \epsilon_{\mathbf{k}_M\mathbf{k}_I}^{n*}\epsilon_{\mathbf{k}_M\mathbf{k}_I}^l g_{\mathbf{k}_M}^c(E) G_{\mathbf{k}_I\sigma,lm}^c(E)
\end{equation}
\end{widetext}
On the l.h.s. we defined the free insulator Green's function. One should note that it includes the interactions through the ISA self-energy. So 'free' does not mean that it does not contain interactions but that it solves the equation of motion for vanishing tunnel coupling. Therefore it can be calculated in thermal equilibrium. This is the essential premise for using the ISA self-energy within this formalism.\\
The equation of motion can be brought into the form of a Dyson equation by defining a tunnel self-energy as follows
\begin{equation}
\Delta_{\mathbf{k}_I,nl}^c(E) = \frac{1}{\hbar^2}\sum_{\substack{M\\M\in L,R}}\sum_{\mathbf{k}_M} \epsilon_{\mathbf{k}_M\mathbf{k}_I}^{n*}\epsilon_{\mathbf{k}_M\mathbf{k}_I}^l g_{\mathbf{k}_M}^c(E)
\end{equation}
With this definition the equation of motion reads
\begin{eqnarray*}
G_{\mathbf{k}_I\sigma,nm}^c(E) &=& g_{\mathbf{k}_I\sigma,nm}^c(E) +\nonumber\\*
&+& \sum_{pq} g_{\mathbf{k}_I\sigma,np}^c(E) \Delta_{\mathbf{k}_I,pq}^c(E) G_{\mathbf{k}_I\sigma,qm}^c(E)
\end{eqnarray*}
The next steps of the calculation are to Fourier transform this equation to time dependence, replace integration over the real time axis by integration over the time contour and then to use the analytic continuation rules again to get the lesser and retarded part of the NEGF. To keep the notation as simple as possible we will use a matrix notation from now on where matrices will be marked by a hat. Then the Dyson equation for the retarded Green's function reads:
\begin{equation}
\hat{G}_{\mathbf{k}_I\sigma}^r(E) = \hat{g}_{\mathbf{k}_I\sigma}^r(E) + \hat{g}_{\mathbf{k}_I\sigma}^r(E) \hat{\Delta}_{\mathbf{k}_I}^r(E) \hat{G}_{\mathbf{k}_I\sigma}^r(E)
\end{equation}
Similarly the lesser Green's function solves the Keldysh equation:
\begin{widetext}
\begin{equation}
\hat{G}_{\mathbf{k}_I\sigma}^<(E) = \left(\hat{1}+\hat{G}_{\mathbf{k}_I\sigma}^r(E) \hat{\Delta}_{\mathbf{k}_I}^r(E)\right) \hat{g}_{\mathbf{k}_I\sigma}^<(E) \left(\hat{1}+\hat{\Delta}_{\mathbf{k}_I}^a(E) \hat{G}_{\mathbf{k}_I}^a(E)\right) + \hat{G}_{\mathbf{k}_I\sigma}^r(E) \hat{\Delta}_{\mathbf{k}_I}^<(E) \hat{G}_{\mathbf{k}_I\sigma}^a(E)
\end{equation}
\end{widetext}
The free lesser insulator Green's function $\hat{g}_{\mathbf{k}_I\sigma}^<(E)$ can be expressed by the free retarded Green's function with the help of the spectral theorem. It doesn't hold in non-equilibrium, but as was pointed out before, the free Green's functions can be calculated in equilibrium. One gets
\begin{equation}
\hat{g}_{\mathbf{k}_I\sigma}^<(E) = -2if_{ins}(E)\mbox{Im} \hat{g}_{\mathbf{k}_I\sigma}^r(E)
\end{equation}
where $f_{ins}(E)$ is the Fermi function in the insulator.

\end{appendix}
\begin{acknowledgments}
We thank S. Mathi Jaya for bringing this topic to our attention. One of us (N.S.) would also like to thank the Friedrich-Ebert-Stiftung for financial support.
\end{acknowledgments}

\end{document}